\documentclass[11pt]{amsart}
\usepackage{amssymb,color}
\numberwithin{equation}{section} \usepackage{amssymb}

\usepackage{graphicx}

\newcommand{\bea}{\begin{eqnarray}}
\newcommand{\eea}{\end{eqnarray}}
\newcommand{\ba}{\begin{array}}
\newcommand{\ea}{\end{array}}
\newcommand{\edc}{\end{document}}
\newcommand{\bc}{\begin{center}}
\newcommand{\ec}{\end{center}}
\newcommand{\be}{\begin{equation}}
\newcommand{\ee}{\end{equation}}

\def\bc{{\mathbb C}}

\def\bn{{\mathbb N}}

\def\m{\mu}

\def\s{\sigma}

\def\w{\omega}
\def\Om{\Omega}

\def\h{{\mathbf{h}}}

\newtheorem{thm}{Theorem}[section]

\newtheorem{prop}[thm]{Proposition}

\theoremstyle{remark}
\newtheorem{rem}{Remark}[section]

\date{\today}
\begin{document}
\title[Gibbs Measures with memory of length 2]{Using New Approaches to obtain Gibbs Measures of Vannimenus model on a Cayley tree}
\author{Hasan Ak\i n}
\address{Department of Mathematics, Faculty of Education,
Zirve University, 27260 Gaziantep, Turkey}

\date{\today }%
\begin{abstract}
 In this paper, we consider Vannimenus model with competing nearest-neighbors and prolonged next-nearest-neighbors interactions  on a Cayley tree. For this model we define Markov random fields with memory of length 2. By using a new approach, we obtain new sets of Gibbs measures of Ising-Vannimenus model on Cayley tree of order 2. We construct the recurrence equations corresponding Ising-Vannimenus  model. We prove the Kolmogorov consistency condition.
We investigate the translation-invariant and periodic non transition-invariant Gibbs measures for the model. We find new sets of Gibbs measures different from the Gibbs measures given in the references \cite{NHSS,FreeMA}.  We show that some of the measures are extreme Gibbs distributions.
\end{abstract}
\maketitle
\section{Introduction}


Two important advantages of using tree models to determine Gibbs measures are that they eliminate the need for approximations and calculations can be carried out to high degrees of accuracy. In addition, models such as Ising and Potts on the Cayley tree can be helpful in discovering additional systems with related properties. As a result, many researchers have employed the Ising and Potts models in conjunction with the Cayley tree (Bethe lattices) \cite{BRZ,PB,bak,BB,BG,GTA,GATTMJ,UGAT}. The Ising model has relevance to physical, chemical, and biological systems \cite{FS,G,IT,Iof}. The Ising model investigated by Vannimenus \cite{Vannimenus} consists of Ising spins $(\sigma=\pm 1)$ on a rooted Cayley tree with a branching ratio of 2 \cite{ART}, in which two coupling constants are present: nearest-neighbour (NN) interactions of strength and next-nearest-neighbour (NNN) interactions. This paper discusses the Ising model corresponding to the Hamiltonian given by Vannimenus \cite{Vannimenus}.

Specifically, researchers used a new method to investigate a rigorous description of Gibbs measures with a memory of length 2 that corresponds to the Ising-Vannimenus Model on the Cayley tree of order 2. Classically, Gibbs states (or measures) only consider finite subsets of $\mathbb{Z}^d$ which are then used to compute various thermodynamic quantities and examine their corresponding limiting behaviour \cite{Bax,D1,FV,G}. For example, Preston studied Gibbs states on a countable tree \cite{Preston}. In addition, Fannes and Verbeure \cite{FV} took into account correlations between $n$ successive lattice points as they studied one-dimensional classical lattice systems with an increasing sequence of subsets of the state space. These states correspond in probability theory to so-called Markov chains with memory of length $n$. Furthermore, Mukhamedov and Akin \cite{FreeMA} proposed a rigorous measure-theoretical approach to investigate Gibbs measures with a memory of length 2 and free energies in the Ising-Vannimenus Model on the Cayley tree. This study further bases its investigation of Gibbs measures on the Markov random field on trees and on recurrent equations following from this theory \cite{ART,AT1,BRZ,BG,GATTMJ,NHSS1,NHSS,GRS,Zachary}. Rozikov, Akin, and Uguz \cite{RAU} analyzed the recurrent equations of a generalized Axial Next-Nearest-Neighbour Ising (ANNNI) model on a Cayley tree and documented critical temperatures and curves, number of phases, and partition function.

This paper also attempts to determine when phase transition occurs. Significant research has determined that a finite graph corresponds to exactly one Gibbs state with potential $F$ for a given potential $F$ and that graphs that are not finite lack this quality, \emph{i.e.}, for some potentials $F$, there may be more than one corresponding Gibbs state with potential $F$ \cite{G,Preston,Sinai}. When there is more than one corresponding Gibbs measure, we say that phase transition occurs for the potential $F$.

Thus, this paper investigates new Gibbs measures with memory of length 2 for generalized ANNNI (Ising-Vannimenus) models on a Cayley tree of order 2. The Markov random field is one of many approaches to studying the equation solutions that describe Gibbs measures for lattice models on Cayley tree, \cite{ART,AT1,BG,NHSS}. This paper uses the Markov random field to achieve the following objectives: construct the recurrence equations corresponding to a generalized ANNNI model; formulate the problem in terms of nonlinear recursion relations along the branches of a Cayley tree; prove the Kolmogorov consistency condition; describe the translation-invariant and periodic (non translation-invariant) Gibbs measures with period 2 for the model; and show that some measures are extreme Gibbs distributions.

This article is organized as follows: Section 2 provides definitions and preliminaries. Section 3 presents a general structure of Gibbs measures with memory of length 2 on a Cayley tree, with functional equations, and proves the Kolmogorov consistency condition. Section 4 describes translation-invariant Gibbs measures corresponding to the associated model \eqref{hm}, demonstrating that some occurrences are extreme. Section 5 constructs periodic non translation-invariant Gibbs measures with period 2. Finally, Section 6 contains concluding remarks and discussion of the consequences of the results.
\section{Preliminaries and Definitions}\label{Preliminaries}
On non-amenable graphs, Gibbs measures depend on boundary conditions \cite{Rozikov}. This paper considers this dependency for Cayley trees, the simplest of graphs. For this paper, let  $\Gamma^k=(V, L, i)$ be the uniform Cayley tree of order $k$ with a root vertex $x^{(0)}\in V$, where each vertex has $k + 1$ neighbors with $V$ as the set of vertices and the set of edges. The notation $i$ represents the incidence function corresponding to each edge $\ell\in L$, with end points $x_1,x_2\in V$. There is a distance $d(x, y)$ on $V$ the length of the minimal point from $x$ to $y$, with the assumed length of 1 for any edge.

We denote the sphere of radius $n$ on $V$ by
$$
W_n=\{x\in V: d(x,x^{(0)})=n \}
$$
and the ball of radius $n$ by
$$
V_n=\{x\in V: d(x,x^{(0)})\leq n \}.
$$
The set of direct successors of $x$ for any $x\in W_n$ is denoted by
$$S(x)=\{y\in W_{n+1}: d(x,y)=1 \}.
$$

The Ising  model with competing nearest-neighbors interactions is defined by the Hamiltonian
\begin{equation*}\label{hm}
H(\sigma)=-J\sum_{<x,y>\subset V}\sigma(x)\sigma(y),
\end{equation*}
where the sum runs over nearest-neighbor vertices $<x,y>$ and the spins $\sigma(x)$ and $\sigma(y)$ take values in the set $\Phi=\{-1,+1\}$.

A finite-dimensional distribution of measure $\mu$  in the volume $V_n$ has been defined by formula
\begin{equation*}\label{mu}
\mu_n(\sigma_n)=\frac{1}{Z_{n}}\exp[-\frac{1}{T}H_n(\sigma)+\sum_{x\in
W_{n}}\sigma(x)h_{x}]
\end{equation*}
with the associated partition function defined as
\begin{equation*}\label{mu}
Z_n=\sum_{\sigma_n \in
\Phi^{V_n}}\exp[-\frac{1}{T}H_n(\sigma)+\sum_{x\in
W_{n}}\sigma(x)h_{x}],
\end{equation*}
where the spin configurations $\sigma_n$ belongs to $\Phi^{V_n}$ and $h=\{h_x\in \mathbb{R},x\in V\}$
is a collection of real numbers that define boundary condition (see \cite{BG,GRRR,GHRR}). Previously, researchers frequently used memory of length 1 over a Cayley tree to study Gibbs measures \cite{BG,GRRR,GHRR}.

The Hamiltonian
\begin{equation}\label{hm}
H(\sigma)=-J_p\sum_{>x,y<}\sigma(x)\sigma(y)-J\sum_{<x,y>}\sigma(x)\sigma(y)
\end{equation}
defines the Ising-Vannimenus model with competing nearest-neighbors and next-nearest-neighbors, with the sum in the first term representing the ranges of all nearest-neighbors. In addition, $J_p,J\in \mathbb{R}$ are coupling constants corresponding to prolonged next-nearest-neighbor and nearest-neighbour potentials.

In \cite{NHSS,FreeMA}, the next generalizations are considered. These authors have defined Gibbs measures or Gibbs states with memory of length 2 for generalized ANNNI models on Cayley trees of order 2 with the following formula:
\begin{equation}\label{mu}
\mu_{\textbf{h}}^{(n)}(\sigma)=\frac{1}{Z_{n}}\exp[-\beta H_n(\sigma)+\sum_{x\in
W_{n-1}}\sum_{y\in S(x)}\sigma(x)\sigma(y)h_{xy,\sigma(x)\sigma(y)}].
\end{equation}
Here, as before, $\beta=\frac{1}{kT}$ and $\sigma_n: x\in V_n\to \sigma_n(x)$ and $Z_n$ corresponds to the following partition function:
$$
Z_n=\sum\limits_{\sigma_n\in \Omega_{V_n}}\exp[-\beta H(\sigma_n)+\sum_{x\in
W_{n-1}}\sum_{y\in S(x)}\sigma(x)\sigma(y)h_{xy,\sigma(x)\sigma(y)}].
$$

We consider increasing subsets of the set of states for one dimensional lattices \cite{FV} as follows: $\mathfrak{G}_1\subset \mathfrak{G}_2\subset... \subset \mathfrak{G}_n\subset...$, where $\mathfrak{G}_n$ is the set of states corresponding to non-trivial correlations between $n$-successive lattice points; $\mathfrak{G}_1$ is the set of mean field states; and $\mathfrak{G}_2$ is the set of Bethe-Peierls states, the latter extending to the so-called Bethe lattices. All these states correspond in probability theory to so-called Markov chains with memory of length $n$.

This paper discusses a new method of defining Markov chains with memory of length 2 using methods described by Fannes and Verbeure \cite{FV}. In \cite{NHSS1,NHSS,FreeMA}, the authors have studied Gibbs measures with memory of length 2 for generalized ANNNI models on a Cayley tree of order 2 by means of a vector valued function 
$$
\textbf{h}:  <x,y> \rightarrow \textbf{h}_{xy}=(h_{xy,++},h_{xy,+-},h_{xy,-+},h_{xy,--})\in \mathbb{R}^4,
$$
where $h_{xy,\sigma(x)\sigma(y)}\in \mathbb{R}$ and $x\in
W_{n-1}, y\in S(x).$

By contrast, this paper assumes that vector valued function is defined by
$$
\textbf{h}:<x,y,z>\rightarrow \textbf{h}_{xyz}=(h_{xyz,+++},h_{xyz,++-},h_{xyz,+-+},h_{xyz,+--},h_{xyz,-++},h_{xyz,-+-},h_{xyz,--+}, h_{xyz,---}),
$$
where $h_{xyz,\sigma(x)\sigma(y)\sigma(z)}\in \mathbb{R}$ and $x\in
W_{n-1}, y,z\in S(x).$

For brevity, we adopt a natural definition for the quantities $h\left(
 \begin{array} {ccc} z, y \\ x
 \end{array} \right)$ as $h_{xyz}$ (see the Fig \ref{fig2}). Finally, we use the function $h_{xyz,\sigma(x)\sigma(y)\sigma(z)}$ to describe the Gibbs measure of any configuration $\left(
 \begin{array} {ccc}\sigma(z), \sigma(y)\\ \sigma(x)
 \end{array}\right)$ that belongs to $\Phi^{V_1}$.


\section{General Structure of Gibbs Measures with Memory of Length 2 on the Cayley tree and Associated Functional Equations}\label{section3}

This section presents the general structure of Gibbs measures with memory of length 2 on the Cayley tree. An arbitrary edge $<x^{(0)},x^{1}>=\ell \in L$ deleted from a Cayley tree $\Gamma^k_1$ and $\Gamma^k_0$ splits into two components: semi-infinite Cayley tree $\Gamma^k_1$ and semi-infinite Cayley tree $\Gamma^k_0$. This paper considers a semi-infinite Cayley tree $\Gamma^k_0$ (see Fig. \ref{fig1}).
\begin{figure} [!htbp]\label{fig1}
\centering
\includegraphics[width=60mm]{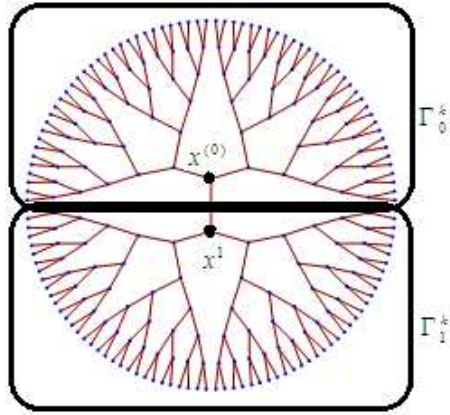}
\caption{Cayley tree of order two, $k=2$.}\label{fig1}
\end{figure}
We define the finite-dimensional Gibbs probability distributions on the configuration space
$$\Omega_{V_n}=\{\sigma_n=\{\sigma(x)=\pm 1, x\in V_n \}\}
$$
at inverse temperature $\beta=\frac{1}{kT}$ by formula
\begin{equation}\label{mu}
\mu_{\textbf{h}}^{(n)}(\sigma)=\frac{1}{Z_{n}}\exp[-\beta H_n(\sigma)+\sum_{x\in
W_{n-1}}\sum_{y,z\in S(x)}\sigma(x)\sigma(y)\sigma(z)h_{xyz,\sigma(x)\sigma(y)\sigma(z)}].
\end{equation}
with the corresponding partition function defined by
$$
Z_n=\sum\limits_{\sigma_n\in \Omega_{V_n}}\exp[-\beta H(\sigma_n)+\sum_{x\in
W_{n-1}}\sum_{y,z\in S(x)}\sigma(z)\sigma(x)\sigma(y) h_{xyz,\sigma(x)\sigma(y)\sigma(z)}].
$$

We will obtain Gibbs measures that differ from previous studies \cite{NHSS,FreeMA}. These new measures consider translation-invariant boundary conditions. We will consider a construction of an infinite volume distribution with given finite-dimensional distributions. More exactly, we will attempt to find a probability measure $\mu$ on $\Omega$  that is compatible with given measures $\mu_{\textbf{h}}^{(n)}$, \emph{i.e.},
\begin{equation}\label{CM}
\m(\s\in\Om: \s|_{V_n}=\s_n)=\m^{(n)}_{\h}(\s_n), \ \ \
\textrm{for all} \ \ \s_n\in\Om_{V_n}, \ n\in\bn.
\end{equation}

The consistency condition for $\mu_{\h}^{n}(\s_n)$, $n\geq 1$ is
\begin{equation}\label{comp}
\sum_{\w\in\Om_{W_n}}\m^{(n)}_{\h}(\s_{n-1}\vee\w)=\m^{(n-1)}_{\h}(\s_{n-1}),
\end{equation}
for any $\s_{n-1}\in\Om_{V_{n-1}}$.
This condition implies the existence of a unique measure $\m_{\h}$ defined on $\Om$  with a required condition \eqref{CM}. Such a measure $\m_{\h}$ is a Gibbs measure with memory of length 2 corresponding to the model. 

We define interaction energy on $V$ with the inner configuration $\s_{n-1}\in V_{n-1}$ and the boundary condition  $\eta \in W_{n}$ as
\begin{eqnarray}\label{ham1}
H_n(\s_{n-1}\vee\eta)
&=&-J\sum\limits_{<x,y>\in V_{n-1}}\sigma(x)\sigma(y) -J
\sum\limits_{x\in W_{n-1}}\sum\limits_{y\in S(x)}\sigma(x)\eta(y)\\\nonumber
&&-J_p\sum\limits_{>x,y<\in V_{n-1}}\sigma(x)\sigma(y) -J_p
\sum\limits_{x\in W_{n-2}}\sum\limits_{z\in S^2(x)}\sigma(x)\eta(z)\\\nonumber
&=&H_n(\s_{n-1})-J\sum\limits_{x\in W_{n-1}}\sum\limits_{y\in S(x)}\sigma(x)\eta(y)-J_p
\sum\limits_{x\in W_{n-2}}\sum\limits_{z\in S^2(x)}\sigma(x)\eta(z).
\end{eqnarray}

The following theorem describes the conditions of quantity $h_{xyz,\sigma(x)\sigma(y)\sigma(z)}$ that guarantee the consistency condition \eqref{comp} of Gibbs measures $\mu_{\h}^{(n)}(\s)$.

\begin{thm}\label{theorem1}
Probability distributions $\mu_{\h}^{(n)}(\s)$, $n=1,2,...,$ in \eqref{mu} are compatible if and only if for any $x,y,z\in V$ ($y,z\in S(x)$) the following equations hold:

\begin{eqnarray}\label{necessary}
e^{h_{xyz,+++}+h_{xyz,-++}}&=&
\frac{\sum\limits_{l,m,n,p\in \{\mp 1\}}a^{(l+m+p+n)}b^{(l+m+n+p)}e^{m lh_{yy_1y_2,+ml}+p n h_{zz_1z_2,+pn]})]}}{\sum\limits_{l,m,n,p\in \{\mp 1\}}a^{(l+m+p+n)}b^{-(l+m+n+p)}e^{[m l h_{yy_1y_2,jml}+ p n h_{zz_1z_2,kpn]})]}}\\\nonumber
e^{h_{xyz,+++}+h_{xyz,---}}&=&
\frac{\sum\limits_{l,m,n,p\in \{\mp 1\}}a^{(l+m+p+n)}b^{(l+m+n+p)}e^{m lh_{yy_1y_2,+ml}+p n h_{zz_1z_2,+pn]})]}}{\sum\limits_{l,m,n,p\in \{\mp 1\}}a^{-(l+m+p+n)}b^{-(l+m+n+p)}e^{[-m l h_{yy_1y_2,-ml}- p n h_{zz_1z_2,-pn}]}}\\\nonumber
e^{h_{xyz,+--}+h_{xyz,---}}&=&
\frac{\sum\limits_{l,m,n,p\in \{\mp 1\}}a^{-(l+m+p+n)}b^{(l+m+n+p)}e^{\left[l m \h_{yy_1y_2,-lm}+n p \h_{zz_1z_2,-np}\right]}}{\sum\limits_{l,m,n,p\in \{\mp 1\}}a^{-(l+m+p+n)}b^{-(l+m+n+p)}e^{[-m l h_{yy_1y_2,-ml}- p n h_{zz_1z_2,-pn}]}}
\end{eqnarray}
where $a=e^{2\beta J}$ and $b=e^{2\beta J_p}$.
\end{thm}
\textbf{Proof:}
Necessity. According to the consistency condition \eqref{comp}, we have
\begin{eqnarray*}
&&L_n\sum\limits_{\eta\in \Omega_{W_{n}}}\exp[-\beta H_n(\s_{n-1}\vee \eta)+\sum\limits_{y,z\in W_{n-1}}\sum\limits_{y_1,y_2\in S(y)}\sum\limits_{z_1,z_2\in S(z)}(B(h,J,J_p)]\\
&=&\exp [-\beta H_n(\s_{n-1})+\sum\limits_{x\in W_{n-2}}\sum\limits_{y,z\in S(x)}\sigma(x)\sigma(y)\sigma(z) h_{xyz,\sigma(x)\sigma(y)\sigma(z)}],
\end{eqnarray*}
where $L_n=\frac{Z_{n-1}}{Z_n}$ and

$$
B(h,J,J_p):=\sigma(y)\eta(y_1)\eta(y_2)h_{yy_1y_2,\sigma(y)\eta(y_1)\eta(y_2)}+\sigma(z)\eta(z_1)\eta(z_2) h_{zz_1z_2,\sigma(z)\eta(z_1)\eta(z_2)}).
$$
Furthermore, equation \eqref{ham1} provides that
\begin{eqnarray*}\label{Kolmogorov1}
&&L_n\sum\limits_{\eta\in \Omega_{W_{n}}}\exp[-\beta H_n(\s_{n-1})-\beta J\sum\limits_{x\in W_{n-1}}\sum\limits_{y\in S(x)}\sigma(x)\eta(y)
-\beta J_p\sum\limits_{x\in W_{n-2}}\sum\limits_{z\in S^2(x)}\sigma(x)\eta(z)\\\nonumber
&&+\sum\limits_{y,z\in W_{n-1}}\sum\limits_{y_1,y_2\in S(y)}\sum\limits_{z_1,z_2\in S(z)}B(h,J,J_p)]\\\nonumber
&=&\exp [-\beta H_n(\s_{n-1})+\sum\limits_{x\in W_{n-2}}\sum\limits_{y,z\in S(x)}\sigma(x)\sigma(y)\sigma(z)h_{xyz,\sigma(x)\sigma(y)\sigma(z)}],
\end{eqnarray*}

%

\begin{eqnarray}\label{Kolmogorov2}
&&L_n\prod\limits_{x\in W_{n-2}}\prod\limits_{y,z\in S(x)}\prod\limits_{y_1,y_2\in S(y)}\prod\limits_{z_1,z_2\in S(z)}\sum\limits_{\eta(y_1),\eta(y_2),\eta(z_1),\eta(z_2)\in \{\mp 1\}}e^{[A(h,J,J_p)]}\\\nonumber
&=&\prod\limits_{x\in W_{n-2}}\prod\limits_{y,z\in S(x)}e^{[\sigma(x)\sigma(y)\sigma(z)h_{xyz,\sigma(x)\sigma(y)\sigma(z)}]},
\end{eqnarray}
where
\begin{eqnarray*}
A(h,J,J_p)&:=&\sigma(y)\eta(y_1)\eta(y_2)h_{yy_1y_2,\sigma(y)\eta(y_1)\eta(y_2)}+\sigma(z)\eta(z_1)\eta(z_2) h_{zz_1z_2,\sigma(z)\eta(z_1)\eta(z_2)})\\
&+&\beta\left[J(\sigma(y)(\eta(y_1)+\eta(y_2))+\sigma(z)(\eta(z_1)+\eta(z_2))+ J_p\sigma(x)(\eta(y_1)+\eta(y_2)+\eta(z_1)+\eta(z_2))\right].
\end{eqnarray*}
\begin{figure} [!htbp]\label{fig2}
\centering
\includegraphics[width=60mm]{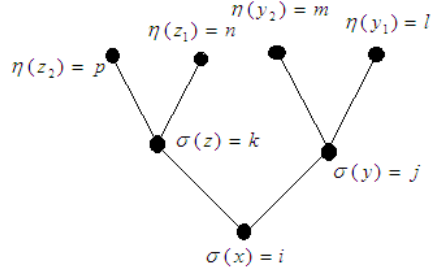}
\caption{Semi-finite Cayley tree of order two with levels 2}\label{fig2}
\end{figure}
Next, let us fix $<x,y>$, $<x,z>$ by rewriting \eqref{Kolmogorov2} for all values of $\sigma(x), \sigma(y), \sigma(z)\in \{-1,+1\}$.

From Figure \ref{fig2} and for brevity we assume $\sigma(x)=i$, $\sigma(y)=j$, $\sigma(z)=k$, $\eta(y_1)=l$, $\eta(y_2)=m$, $\eta(z_1)=n$, $\eta(z_2)=p$,
where $i,j,k,l,m,n,p\in \{-1,+1\}$.\\
Then from \eqref{Kolmogorov2}, we have
\begin{eqnarray}\label{exponent}\nonumber
e^{ijk h_{xyz,ijk}}&=&L_2\sum\limits_{l,m,n,p\in \{\pm 1\}}e^{[\beta(J(j(l+m)+k(p+n))+iJ_p(l+m+n+p))+jm l h_{yy_1y_2,jml}+k p n h_{zz_1z_2,kpn]})]}.\\
\end{eqnarray}
Because of the arbitrariness in the choice of the origin, we must have $h_{[+;-+]}=h_{[+;+-]}$, $h_{[-;-+]}=h_{[-;+-]}$.
Thus, we obtain
\begin{eqnarray*}
D_1=e^{h_{xyz,+++}}&=&L_2\frac{\left(2 a b+e^{h_{[+;--]}+ h_{[+;-+]}}+(a b)^{2}e^{h_{[+;-+]}+ h_{[+;++]}}\right)^2}{(a b)^{2}e^{2h_{[+;-+]}}}\\
D_2=e^{h_{xyz,+--}}&=&L_2\frac{\left(b^{2}e^{h_{[---]}}+a^{2}e^{h_{[-++]}}+2 (a b)e^{h_{[---]}+h_{[--+]}+h_{[-++]}}\right)^2}{(a b)^{2}e^{2\h_{[---]}+2\h_{[-++]}}}\\
\end{eqnarray*}
\begin{eqnarray*}
D_3=e^{-h_{xyz,-++}}&=&L_2\frac{(2(a b) + b^{2}e^{h_{[+--]}+h_{[+-+]}} +a^{2}e^{h_{[+-+]}+h_{[+++]}})^2}{(a b)^{2}e^{2h_{[+-+]}}}\\
D_4=e^{-h_{xyz,---}}&=&L_2\frac{\left(e^{h_{[---]}}+(a b)^{2}e^{h_{[-++]}}+2 (a b)e^{h_{[---]}+h_{[--+]}+h_{[-++]}}\right)^2}{(a b)^{2}e^{2h_{[---]}+2h_{[-++]}}}.
\end{eqnarray*}
Finally we have
\begin{equation*}\label{Kolmogorov3}
\frac{D_1}{D_3}:=e^{h_{xyz,+++}+h_{xyz,-++}}=
\frac{\sum\limits_{l,m,n,p\in \{\mp 1\}}e^{[\beta(J((l+m)+(p+n))+J_p(l+m+n+p))+
m l \ h_{yy_1y_2,+ml}+p n \ h_{zz_1z_2,+pn]})]}}{\sum\limits_{l,m,n,p\in \{\mp 1\}}e^{[\beta(J((l+m)+k(p+n))-J_p(l+m+n+p))+
m l \ h_{yy_1y_2,jml}+ p n \ h_{zz_1z_2,kpn]})]}}
\end{equation*}
\begin{equation*}\label{Kolmogorov4}
\frac{D_1}{D_4}:=e^{ h_{xyz,+++}+h_{xyz,---}}=
\frac{\sum\limits_{l,m,n,p\in \{\mp 1\}}e^{\left[l m \ h_{yy_1y_2,+lm}+n p \ h_{zz_1z_2,+np}+\beta (J(l+m+n+p)+ J_p(l+m+n+p))\right]}}{\sum\limits_{l,m,n,p\in \{\mp 1\}}e^{[-m l \ h_{yy_1y_2,-ml}- p n \ h_{zz_1z_2,-pn}+\beta(-J(l+m+p+n)-J_p(l+m+n+p))]}}
\end{equation*}
\begin{equation*}\label{Kolmogorov4a}
\frac{D_2}{D_3}:=e^{h_{xyz,+--}+h_{xyz,---}}=
\frac{\sum\limits_{l,m,n,p\in \{\mp 1\}}e^{\left[l m\ h_{yy_1y_2,-lm}+n p\ h_{zz_1z_2,-np}-\beta (J(l+m+n+p)+ J_p(l+m+n+p))\right]}}{\sum\limits_{l,m,n,p\in \{\mp 1\}}e^{[-m l\ h_{yy_1y_2,-ml}- p n \ h_{zz_1z_2,-pn}+\beta(-J(l+m+p+n)-J_p(l+m+n+p))]}}
\end{equation*}

These equations imply the desired equations. This completes the proof.

Sufficiency: Now assume that the system of equations in \eqref{necessary} is valid, then from \cite[Lemma 2.4]{MDA} we derive

\begin{eqnarray*}
e^{\sigma(x)\sigma(y)\sigma(z)h_{xyz,\sigma(x)\sigma(y)\sigma(z)}}D(x,y,z)
&=&\prod\limits_{y_1,y_2\in S(y)}\prod\limits_{z_1,z_2\in S(z)}\sum\limits_{\eta(y_1),\eta(y_2),\eta(z_1),\eta(z_2)\in \{\mp 1\}}e^{A(h,J,J_p)}.
\end{eqnarray*}
From the previous equation, one can conclude
\begin{eqnarray}\label{Kolmogorov5}
&&\prod\limits_{x\in W_{n-2}}\prod\limits_{y,z\in S(x)}D(x,y,z)e^{\sigma(x)\sigma(y)\sigma(z)h_{xyz,\sigma(x)\sigma(y)\sigma(z)}}\\\nonumber
&=&\prod\limits_{x\in W_{n-2}}\prod\limits_{y,z\in S(x)}\prod\limits_{y_1,y_2\in S(y)}\prod\limits_{z_1,z_2\in S(z)}\left[\sum\limits_{\eta(y_1),\eta(y_2),\eta(z_1),\eta(z_2)\in \{\mp 1\}}e^{A(h,J,J_p)}\right].
\end{eqnarray}
Multiply both sides of the equation \eqref{Kolmogorov5} by $e^{-\beta H_{n-1}(\sigma)}$ results in
$$
U_n=\prod\limits_{x\in W_{n-2}}\prod\limits_{y,z\in S(x)}D(x,y,z),
$$
From \eqref{Kolmogorov5} we have
\begin{eqnarray*}
&&U_{n-1}e^{\{-\beta H_{n-1}(\sigma)+\sum\limits_{x\in W_{n-2}}\sum\limits_{y,z \in S(x)}\sigma(x)\sigma(y)\sigma(z) h_{xyz,\sigma(x)\sigma(y)\sigma(z)}\}}\\
&=&\prod\limits_{x\in W_{n-2}}\prod\limits_{y\in S(x)}\prod\limits_{y_1,y_2\in S(y)}\prod\limits_{z_1,z_2\in S(z)}e^{-\beta H_{n-1}(\sigma)}\left[\sum\limits_{\eta(y_1),\eta(y_2),\eta(z_1),\eta(z_2)\in \{\mp 1\}}e^{A(h,J,J_p)}\right].
\end{eqnarray*}
Therefore, we can conclude
$$
U_{n-1}Z_{n-1}\mu_{n-1}(\sigma)=\sum\limits_{\eta}e^{\left(-\beta H_{n}(\sigma\vee\eta)+\sum\limits_{x\in W_{n-2}}\sum\limits_{y,z\in S(x)}\sigma(x)\sigma(y)\sigma(z)h_{xyz,\sigma(x)\sigma(y)\sigma(z)}\right)}
$$
\begin{eqnarray}\label{eq4}
U_{n-1}Z_{n-1}\mu_{n-1}(\sigma)=Z_{n}\sum\limits_{\eta}\mu_{n}(\sigma\vee\eta).
\end{eqnarray}
As $\mu_n$ ($n\geq 1$) is a probability  measures, i.e.
$$
\sum\limits_{\sigma\in \{-1,+1\}^{V_{n-1}}}\mu_{n-1}(\sigma)=\sum\limits_{\sigma\in \{-1,+1\}^{V_{n-1}}}\sum\limits_{\eta \in \{-1,+1\}^{W_{n}}}\mu_{n}(\sigma\vee\eta)=1
$$
so from these equalities and the equation \eqref{eq4} we have $Z_{n}=U_{n-1}Z_{n-1}$.

Thus, as a result of equation \eqref{eq4}, we can imply that \eqref{comp} holds.
\section{Translation-invariant Gibbs measures with memory of length 2}
This section establishes the existence of Gibbs measures by analyzing the equations \eqref{necessary}. Recall that a function $\h = \{h_{xyz,\sigma(x)\sigma(y)\sigma(z)}: \sigma(x),\sigma(y),\sigma(z)\in \{-1,+1\} \}$ is considered translation-invariant if $h_{xyz,ijk}= h_{uvw,ijk}$ for all $<x,y>, <x,z>, <u,v>,$ $ <u, w> \in L$ and $i,j,k\in \{-1,+1\}$. A translation-invariant Gibbs measure is defined as a measure, $\mu_{\h}$, corresponding to a translation-invariant function $\h$.

\subsection{Basic Equations}
Assume that $a=e^{2\beta J}$ and $b=e^{2\beta J_p}$. From \eqref{exponent}, we obtain the following six equations:
\begin{eqnarray*}
e^{h_{[+;++]}}&=&\frac{\left(2 (ab)+e^{h_{[+;--]}+h_{[+;-+]}}+(ab)^{2}e^{h_{[+;-+]}+h_{[+;++]}}\right)^2}{(ab)^{2}e^{2h_{[+;-+]}}}\\
e^{-h_{[+;-+]}}&=&\frac{(2 (ab)+e^{h_{[+;-+]}}(e^{h_{[+;--]}}+(ab)^{2}e^{h_{[+;++]}}))(b^{2}e^{h_{[-;--]}}+e^{h_{[-;++]}}(a^{2}+2 (ab)e^{h_{[-;--]}+h_{[-;-+]}})}{(ab)^{2}e^{h_{[-;--]}+h_{[-;++]}+h_{[+;-+]}}}\\
e^{h_{[+;--]}}&=&\frac{\left(b^{2}e^{h[-;--]} +a^{4}e^{h[-;++]} +2 (ab)e^{h[-;--]+h[-;-+]+h[-;++]}\right)^2}{(ab)^{2}e^{2(h[-;--]+ h[-;++])}}\\
e^{-h_{[-;++]}}&=&\frac{\left(2 (ab)+b^{2}e^{h[+;--]+h[+;-+]}+a^{2}e^{h[+;-+]+h[+;++]}\right)^2}{(ab)^{2}e^{2h[+;-+]}}\\
e^{h_{[-;-+]}}&=&\frac{\left(2(ab)+e^{h_{[+;-+]}}(b^{2}e^{h_{[+;--]}}+a^{2}e^{h_{[+;++]}})\right)\left(e^{h_{[-;--]}}+e^{h_{[-;++]}}((ab)^{2}+2 (ab)e^{h_{[-;--]}+h_{[-;-+]}})\right)}{(ab)^{2}e^{h_{[-;--]}+h_{[-;++]}+h_{[+;-+]}}}\\
e^{-h_{[-;--]}}&=&\frac{\left(e^{h[-;--]}+(ab)^{2} e^{h[-;++]}+2 (ab)e^{h[-;--]+ h[-;-+]+h[-;++]}\right)^2}{(ab)^{2}e^{2h[-;--]+2h[-;++]}}
\end{eqnarray*}
For convenience, we will use a shorter notation for the recurrence system \cite{Vannimenus}:\\
$h_{xyz,+++}=h_{[+;++]}=\ln (u_1)$, $h_{xyz,+-+}=h_{xyz,++-}=h_{[+;-+]}=\ln (u_2)$, $h_{xyz,+--}=h_{[+;--]}=\ln (u_3)$, $h_{xyz,-++}=h_{[-;++]}=\ln (u_4)$, $h_{xyz,-+-}=h_{xyz,--+}=h_{[-;+-]}=\ln (u_5)$, $h_{xyz,---}=h_{[-;--]}=\ln (u_6)$.\\
Next, through direct enumeration, we have
\begin{eqnarray*}\label{recurrence1}
u_1&=&\frac{\left(2 (ab)+u_2 u_3+(ab)^{2}u_1u_2\right)^2}{(ab)^{2}u_2^2}\\
u_2^{-1}&=&\frac{\left(2 (ab)+u_2 u_3+(ab)^{2}u_1u_2\right)(b^{2}u_6 + a^{2}u_4+2 (ab)u_4 u_5u_6)}{(ab)^{2}u_2u_4u_6}\\
u_3&=&\frac{\left(b^{2}u_6 + a^{2}u_4+2 (ab)u_4 u_5u_6\right)^2}{(ab)^{2}u_4^2u_6^2}\\
u_4^{-1}&=&\frac{\left(2 (ab)+b^{2}u_2u_3+a^{2}u_1u_2\right)^2}{(ab)^{2}u_2^{2}}\\
u_5&=&\frac{\left(2 (ab)+b^{2}u_2u_3+a^{2}u_1u_2\right)\left(u_6+(ab)^{2}u_4+2 (ab)u_4u_5u_6\right)}{(ab)^{2}u_2u_4u_6}\\
u_6^{-1}&=&\frac{\left(u_6+(ab)^{2}u_4+2 (ab)u_4u_5u_6\right)^2}{(ab)^{2}u_4^2u_6^2}.\\\nonumber
\end{eqnarray*}
%
%
%
Note that $u_2^{-2}=u_1 u_3$ and $u_5^{2}=\frac{1}{u_4 u_6}$; therefore, we obtain only 4 variables. When we consider new variables such as $u_i=\sqrt{v_i}$, the new recurrence system can be expressed in a simpler form:
\begin{eqnarray}\label{recurrenceMain1}
v_1&=&(ab)^{-1}(v_3+(ab)v_1)^2\\\label{recurrenceMain2}
v_3&=&(ab)^{-1}\left(\frac{b}{v_4}+\frac{a}{v_6}\right)^2\\\label{recurrenceMain3}
v_4^{-1}&=&(ab)^{-1}\left(av_1+bv_3\right)^2\\\label{recurrenceMain4}
v_6^{-1}&=&(ab)^{-1}\left(\frac{1}{v_4}+\frac{(ab)}{v_6}\right)^2.\\\nonumber
\end{eqnarray}
Creating conditions favourable to the occurrence of phase transition depends in part on finding a so-called critical temperature. Note that the equations above describe the fixed points of equation \eqref{comp}, which satisfies the consistency condition. When there is more than 1 solution for equations \eqref{necessary}, then more than 1 translation-invariant Gibbs measure corresponds to those solutions. When equations \eqref{recurrence3} and \eqref{recurrence4} have more than 1 positive solution, a phase transition occurs for model \eqref{hm}. This possible non-uniqueness corresponds in the language of statistical mechanics to the phenomenon of phase transition \cite{Preston}. Phase transitions usually occur at low temperatures. Finding an exact value for $T_c$, where $T_c$ is the critical value of temperature, can enable the creation of conditions in which a phase transition occurs for all $T$. Solving models for $T_c$ will lead to finding the exact value of the critical temperatures.

The number of the solutions of the Equations \eqref{recurrence5}, \eqref{recurrence12} and \eqref{recurrence15} naturally depends on the parameter $\beta=1/kT$. Thus, we will find positive fixed points of the nonlinear dynamical systems \eqref{recurrence6a}, \eqref{recurrence12a} and \eqref{recurrence15}, respectively.


For the second and the third cases, we will use same method, \emph{i.e.}, obtain the fixed points of the recurrences equations.
\subsection{First Case}
From the equations \eqref{recurrenceMain1} through \eqref{recurrenceMain4}, assume that $v_1=v_4$ and $v_3=v_6$, \emph{ i.e.} $h_1=h_{[+;++]}=h_{[-;++]}$ and $h_2=h_{[+;--]}=h_{[-;--]}$. \\
 Next, we obtain
\begin{eqnarray}\label{recurrence3}
v_1^{2}&=&\left(\frac{v_3+(a b)v_1}{bv_3+a v_1}\right)^2
\end{eqnarray}

\begin{eqnarray}\label{recurrence4}
v_3^{2}&=&\left(\frac{b v_3+a v_1}{v_3+abv_1}\right)^2\\\nonumber
\end{eqnarray}
From the equations \eqref{recurrence3} and \eqref{recurrence4}, one can obtain the following nonlinear dynamical system:
\begin{equation}\label{recurrence5}
v=\frac{a b v^2+1}{b+a v^2}\\
\end{equation}
The fixed points $\h = F(\h)$ of equation \eqref{recurrence5}, where $\h = (h_1, h_2)$, describe translation-invariant Gibbs measures with memory of length 2 for the Ising-Vannimenus model. Let us now investigate the fixed points of the dynamic system \eqref{recurrence5}, i.e., $v = f(v)$, where
\begin{eqnarray}\label{recurrence6a}
f(v):=\frac{a b v^2+1}{b+a v^2}.
\end{eqnarray}
If we define $f: \mathbb{R}^{+}\rightarrow \mathbb{R}^{+}$
then $f$ is bounded and thus the curve $y = f(v)$ must intersect
the line $y =m v.$ Therefore, this construction provides one element of
a new set of Gibbs measures with memory of length 2, corresponding to the model  \eqref{hm} for any $v\in \mathbb{R}^{+}$.
\begin{prop}
The equation
$$
v=\left(\frac{a b v^2+1}{b+a v^2}\right)
$$
(with $v \geq0, a > 0, b > 0$) has one solution if  $b<1$.
If $b>\sqrt{3}$ then there exists $\eta_1(a,b)$,
$\eta_2(a,b)$ with $0<\eta_1(a,b)<\eta_2(a,b)$ such that equation \eqref{recurrence6a} has 3 solutions if $\eta_1(a,b)<m<\eta_2(a,b)$ and has 2 solutions if either $\eta_1(a,d)=m$ or $\eta_2(a,b)=m$, where
\begin{eqnarray*}
\eta_1 (a,b)&=&-\frac{2 a b^2 \left(-9-2 a b+8 b^2-b^4+(b^2-3)\sqrt{9-10 b^2+b^4}\right)}{\left(3-b^2+\sqrt{9-10 b^2+b^4}\right) \left(-9+8 b^2-2 a b^3-b^4+(b^2-3)\sqrt{9-10 b^2+b^4}\right)}\\
\eta_2 (a,b)&=&\frac{2 a b^2 \left(9+2 a b-8 b^2+b^4+(b^2-3)\sqrt{9-10 b^2+b^4}\right)}{\left(-3+b^2+\sqrt{9-10 b^2+b^4}\right) \left(9-8 b^2+2 a b^3+b^4+(b^2-3)\sqrt{9-10 b^2+b^4}\right)}.
\end{eqnarray*}
\end{prop}
\textbf{Proof:}
Let
$$
f(v)=\left(\frac{a b v^2+1}{b+a v^2}\right).
$$
Then, taking the first and the second derivatives of the function $f$, we have
$$
f'(v)=\frac{2a (b^2-1)v}{\left(b+a v^2\right)^2}
$$
$$
f''(v)=-\frac{2 a (b^2-1) \left(3 a v^2-b\right)}{\left(b+a v^2\right)^3}.
$$
If $b<1$ (with $v \geq0$) then $f$ is decreasing and there can only be one solution of $f(v) = v.$
Thus, we can restrict ourselves to the case in which $b>1.$  In this case, $f$ is convex for $v < \sqrt{\frac{b}{3a}}$
and is concave for $v > \sqrt{\frac{b}{3a}}$. As a result, there are at most 3 solutions for $f(v) = v.$
According to Preston \cite{Preston} in Proposition 10.7, there can be more than one solution if and only if there is more than
one solution to $vf''(v) = f(v)$, which is the same as
\begin{equation}\label{root1}
a^2 b v^4-a\left(b^2-3 \right) v^2+b=0.
\end{equation}
With some elementary analysis, we have
$$
b>\sqrt{3},\ \ \ \Delta=a^2 (b^2-9) (b^2-1)>0
$$
and
$$
v_1^2=\frac{ (b^2-3)-\sqrt{(b^2-9) (b^2-1)}}{2 ab},\ \ \ \ \ \ \ v_2^2=\frac{ (b^2-3)+ \sqrt{(b^2-9) (b^2-1)}}{2 ab}.
$$
where $b>3$ due to $b >1$. Then $f'(v_1) < 1$ and $f'(v_2) > 1$. That is, $f(v_1) < v_1$ and $f(v_2) > v_2$, if
$\eta_1(a,b) < 1 <\eta_2(a,b)$.

Thus, the proof is readily completed.
\begin{figure} [!htbp]\label{fig1Cregion}
\centering
\includegraphics[width=60mm]{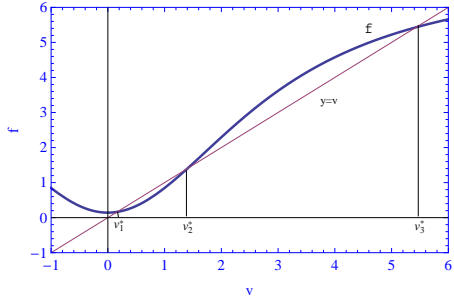}\ \ \ \ \ \ \ \ \ \ \ \ \label{fig1c}
\caption{Three positive roots of the equation \eqref{recurrence5} for $a = 0.8, b = 7$  
.}\label{fig1Cregion}
\end{figure}
\begin{figure} [!htbp]\label{fig1c3D}
\centering
\includegraphics[width=60mm]{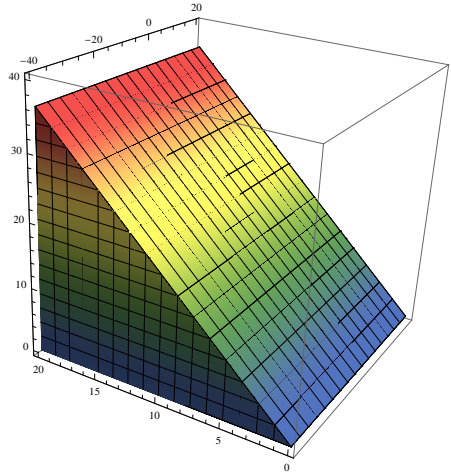}
\caption{Three-dimensional region for the positive three roots of the equation \eqref{recurrence6a}. For all $(J,J_p,T)$ inside the surface, a phase transition occurs ($J\in (-40, 20), J_p\in (0, 20), T\in(0.00012, 40)$.}\label{fig1c3D}
\end{figure}

Figure \ref{fig1Cregion} illustrates that for all $v\in (v_2^{*},v_3^{*})$, $\lim\limits_{n\rightarrow \infty}f^n(v)=v_3^{*}.$ Similarly, for all $v\in (v_1^{*},v_2^{*})$, $\lim\limits_{n\rightarrow \infty}f^n(v)=v_1^{*}.$
Therefore, the fixed points $v_1^{*}$ and $v_3^{*}$ are stable and $v_2^{*}$ is unstable.
\begin{rem}
We can conclude that the Gibbs measures  $\mu_1^{*}$ and $\mu_3^{*}$ corresponding to the stable fixed points $v_1^{*}$
and $v_3^{*}$ are extreme Gibbs distributions (for details \cite{NHSS,Iof}).
\end{rem}
\subsection{Second Case}
From the equations \eqref{recurrenceMain1} through \eqref{recurrenceMain4}, assume that $x= v_1=\frac{1}{v_3}$ and $y=v_6=\frac{1}{v_4}$, i.e. $h_1=h_{[+;++]}=-h_{[+;--]}$ and $h_2=h_{[-;--]}=-h_{[-;++]}$. As a result, we get
\begin{eqnarray}\label{recurrence9}
x^2=
\frac{\left(1+a b x^2\right) y}{\left(a+b y^2\right)}
\end{eqnarray}
Secondly, we can obtain
\begin{eqnarray}\label{recurrence10}
x=
\frac{\left(b+a x^2\right)}{\left(a b+y^2\right)}.
\end{eqnarray}
From \eqref{recurrence9} and \eqref{recurrence10}, we have

\begin{eqnarray}\label{recurrence11}
x^2=
\left(\frac{(b+\frac{a y}{a-a b y+b y^2})}{\left(a b+y^2\right)}\right)^2.
\end{eqnarray}
From \eqref{recurrence9} and \eqref{recurrence11}, we obtain
\begin{eqnarray}\label{recurrence12}
y
=\frac{\left(a b-a\left(b^2-1\right) y+b^2y^2\right)^2}{\left(a b+y^2\right)^2 \left(a-a b y+b y^2\right)}
:=g(y).
\end{eqnarray}
By investigating the fixed points of the rational function $g(y)$,  the equation \eqref{recurrence12} has the following form:
\begin{eqnarray}\label{recurrence12a}
y=\frac{\left(a b-a\left(b^2-1\right) y+b^2y^2\right)^2}{\left(a b+y^2\right)^2 \left(a-a b y+b y^2\right)},
\end{eqnarray}
which is equivalent to the following equation:
\begin{equation}\label{recurrence13b}
(-a+a b y-b y^2+y^3)(a b^2-a b( a b+ b^2-2) y+(a+b^3) y^2-b(a -b) y^3+b y^4)=0.
\end{equation}
We use Descartes' Rule of Signs to find the zeroes of a polynomial.  Thus, we determine the number of real solutions to the equation \eqref{recurrence13b}. Furthermore, if $a b+ b^2-2>0$ and $ a>b$ there are seven sign changes in the positive case. Seven is the maximum possible number of positive zeroes 
for the polynomial in the equation \eqref{recurrence13b}.

Next we will examine $p_7(-y)$, which is the negative case of the polynomial:
$$
p_7(-y):=(-a-a b y-b y^2-y^3)(a b^2+a b( a b+ b^2-2) y+(a+b^3) y^2+b(a -b) y^3+b y^4).
$$
By counting the number of sign changes, we can determine that if $a b+ b^2-2>0$ and $a>b$, then there is no sign change in this negative case, so there are no negative roots. Therefore, if $a b+ b^2-2>0$ and $ a>b$, there are 7, 5, 3, or 1 positive roots, and no negative roots.
Also, it is obvious that $\lim\limits_{y\rightarrow \infty}g(y)=0$ any $g(0)=\frac{1}{a}>0$, there is at least 1 positive root such that $g(y)=y$.
\begin{figure} [!htbp]\label{fig4}
\centering
\includegraphics[width=70mm]{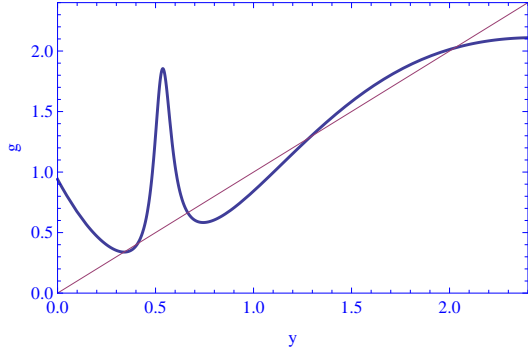}
\caption{The positive roots of the equation \eqref{recurrence12} for $a = 1.064$ and $b = 3.72$}\label{fig4}
\end{figure}
\begin{figure} [!htbp]\label{fig5}
\centering
\includegraphics[width=60mm]{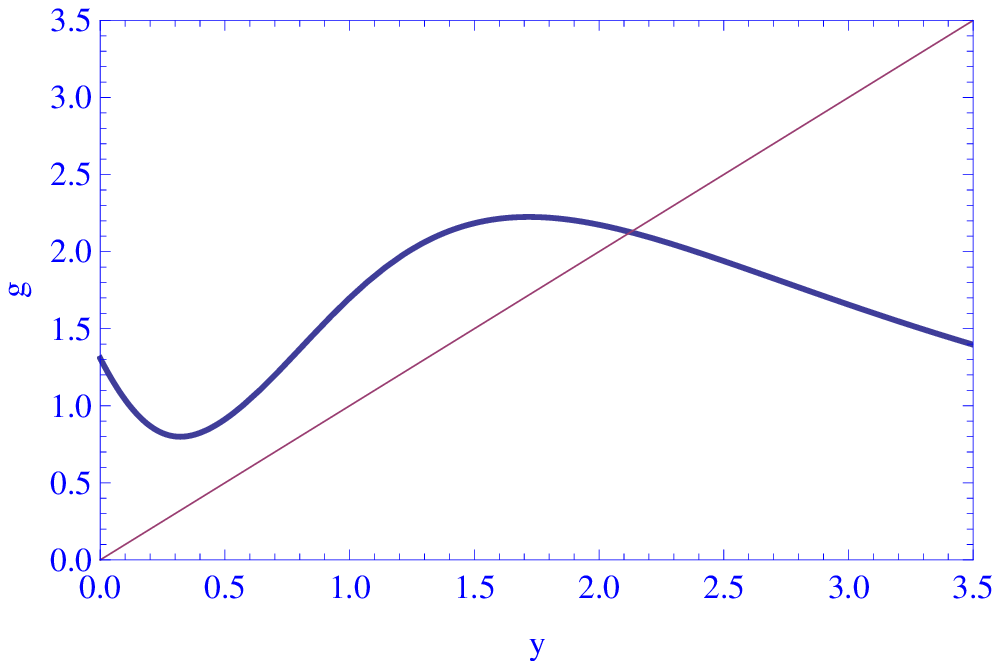}\ \ \ \ \ \ \ \label{fig5a}
\includegraphics[width=60mm]{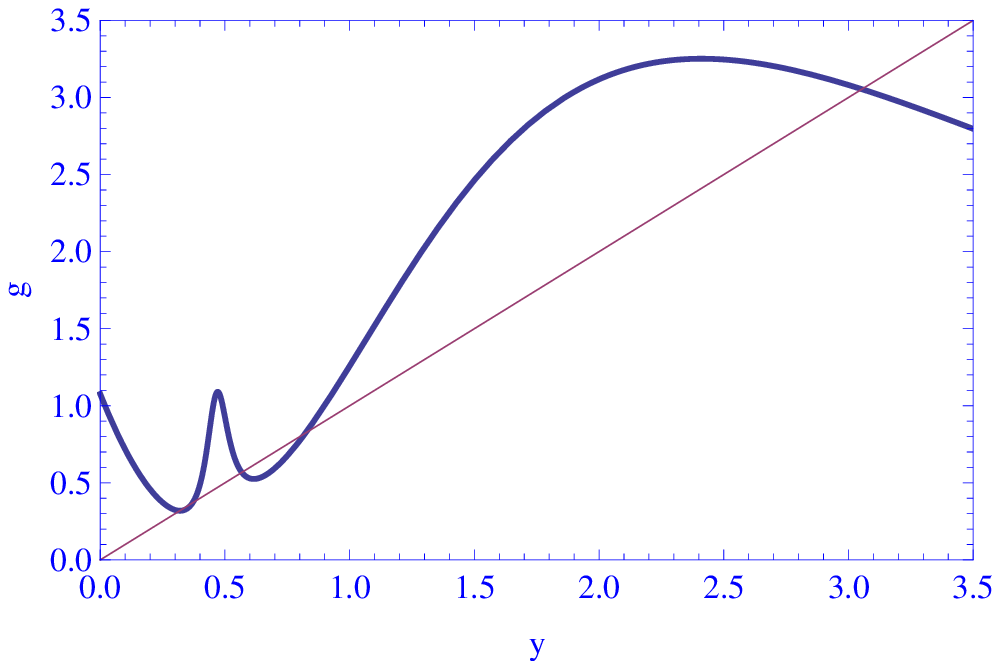} \label{fig5b}
\caption{(a, left) There exists a positive root of the equation \eqref{recurrence12} for $J=-1.2, T = 8.95, J_p = 5$ and $m=1$,
(b, right) There exist five positive of the equation \eqref{recurrence12} for $J = -0.33, T = 8.99, J_p = 65$ and $m=1$.}
\label{fig5}
\end{figure}
Using another approach,
$$
g'(y)=\frac{\left(a b-a(b^2-1) y+b^2 y^2\right)A(y)}{\left(a b+y^2\right)^3 \left( a-a b y+b y^2\right)^2},
$$
where
\begin{eqnarray*}
A(y)&=&a^3 b \left(2-b^2\right)+a^2 b \left(-4-a b+2 b^2+a b^3\right) y-a^2 \left(-2+b^2\right) \left(-1+3 b^2\right)y^2\\
&&+a b \left(3 a-6 b-3 a b^2+2 b^3\right)y^3+a b \left(-4+5 b^2\right) y^4-2 b^3 y^5
\end{eqnarray*}
On the other hand, equation \eqref{recurrence12} has more than one solution if and only if there is more than one solution of the
$yg'(y)=g(y)$, which is the same as
\begin{eqnarray}\label{recurrence12teg}
(a b-a(b^2-1) y+b^2 y^2)(A_1+B_1 y+C_1 y^2+D_1 y^3+E_1 y^4+F_1 y^5+b^3 y^6)=0,
\end{eqnarray}
where
\begin{eqnarray*}
A_1&=&-2 a^3 b+a^3 b^2+a^3 b^3\\
B_1&=&(4 a^2 b+a^3 b+a^3 b^2-2 a^2 b^3-2 a^3 b^3-a^3 b^4)\\
C_1&=&(2 a^2+a^2 b-7 a^2 b^2-a^3 b^2+2 a^2 b^3+3 a^2 b^4+a^3 b^4)\\
D_1&=&(a^2-3 a^2 b+6 a b^2-a^2 b^2+3 a^2 b^3-2 a b^4-2 a^2 b^4)\\
E_1&=&(4 a b-a^2 b+2 a b^2-5 a b^3+a^2 b^3+a b^4)\\
F_1&=&(a b+2 b^3-2 a b^3).
\end{eqnarray*}

Standard calculus arguments describe the region where there are at most 7 positive fixed points,
\emph{i.e.}, the region where phase transition occurs. In this case, also for parameters $a$ and $b$ in the phase transition region, equation \eqref{recurrence12} has at most 7 positive roots, at least 2 of them stable, the others potentially unstable. In Figure \ref{fig4}, for $a = 1.064$ and $b = 3.72$, there are 5 positive fixed points, at least 2 of them stable, the others unstable. Elementary analysis allows us to obtain positive fixed points of function \eqref{recurrence12a} by finding real positive roots of equation \eqref{recurrence12}. Thus, we obtain a polynomial equation of degree 7. Previously documented analysis has solved these equations, which we will not show here due to the complicated nature of formulas and coefficients \cite{Wolfram}. Nonetheless, we have manipulated the polynomial equation $p_7(y)$ via Mathematica \cite{Wolfram}.

In Figure \ref{fig5} (a), there exists only 1 positive root of equation  \eqref{recurrence12} for $J = -1.2, T = 8.95, J_p = 5$. In Figure \ref{fig5} (b), there exist 5 positive roots of equation \eqref{recurrence12} for $J = -0.33, T = 8.99, J_p = 6.5$; therefore, for $J = -0.33, T = 8.99, J_p = 6.5$, we have 5 translation-invariant Gibbs measures corresponding to the fixed points. In this case, a phase transition occurs.

Therefore, in the Second Case, there is a critical temperature $T_c > 0$ such that for $T < T_c$ this system of equations has 5 solutions: $h_1^{*}; h_2^{*};h_3^{*};h_4^{*};h_5^{*}.$  We denote the Gibbs measure that corresponds to the root $h_1^{*}$ (and respectively $h_2^{*};h_3^{*};h_4^{*};h_5^{*}$) by $\mu^{(1)}$ (and respectively $\mu^{(2)}$,$\mu^{(3)}$,$\mu^{(4)}$,$\mu^{(5)}$).

\begin{rem}
Note that the stable roots describe extreme Gibbs distributions \cite{NHSS1,NHSS,Iof}.
\end{rem}
\subsection{Third Case}

Finally, from equations \eqref{recurrenceMain1} through \eqref{recurrenceMain4}, assume that $v_1=v_6$ and $v_3=v_4$,\emph{ i.e.} $h_1=h_{[+;++]}=h_{[-;--]}$ and $h_2=h_{[+;--]}=h_{[-;++]}$. Then, we get
\begin{eqnarray}\label{recurrence7}
v_3&=&\left(\frac{v_3(a b)+v_1}{v_3+(a b)v_1}\right)\\\label{recurrence8}
v_3^{2}&=&\left(\frac{b v_1+a v_3}{v_1(a v_1+b v_3)}\right)\\\nonumber
\end{eqnarray}
For simplicity, we assume $v_1=y$ and $v_3=x$.
From \eqref{recurrence7}, we substitute $y=\frac{x(x-a b)}{ (1- a b x)}$. Then from equation \eqref{recurrence8} we have
$$
x^2=\frac{-a+a b^2+\left(-b+2 a^2 b-a^2 b^3\right) x+\left(a b^2-a^3 b^2\right) x^2}{\left(a b^2-a^3 b^2\right) x+\left(-b+2 a^2 b-a^2 b^3\right) x^2
+\left(-a+a b^2\right)x^3}.
$$
As a result, we have
\begin{eqnarray}\label{recurrence15}
x=\sqrt{\frac{-a+a b^2+\left(-b+2 a^2 b-a^2 b^3\right) x+\left(a b^2-a^3 b^2\right) x^2}{\left(a b^2-a^3 b^2\right) x+\left(-b+2 a^2 b-a^2 b^3\right) x^2+\left(-a+a b^2\right) x^3}}:=F(x)
\end{eqnarray}
From the equation \eqref{recurrence15} we have
\begin{eqnarray}\label{recurrence16}
(A x^4-B x^3+C x^2-B x+A)(x-1)=0,
\end{eqnarray}
where
$$
A=a(b^2-1);\ \ B=(a+b-2 a^2 b-a b^2+a^2 b^3);\ \ \ C=(-a-b+2 a^2 b+2 a b^2-a^3 b^2-a^2 b^3).
$$
This proof makes clear that $x=1$ is one of the positive roots of equation \eqref{recurrence15}. Next, we will find the other roots.

By using Descartes' Rule of Signs to find the zeroes of a polynomial, we can determine the number of real solutions to equation \eqref{recurrence16}.

Clearly, if $A>0,\ B>0$ and $C>0$ for the polynomial $p_4(x)=(A x^4-B x^3+C x^2-B x+A)$ then there are 4 sign changes in the positive case, and a maximum of 4 positive zeroes ($y$-intercepts) for the polynomial $p_4(x)$.
Next, we will examine the negative polynomial ($p_4(-x)$):
$$
p_x(-x):=(A x^4+B x^3+C x^2+B x+A).
$$
We can count the number of sign changes. If $A>0,\ B>0$ and $C>0$, then there is no sign change in this negative case, so there are no negative roots. Therefore, if $A>0,\ B>0$ and $C>0$, there are either 4, 2, or 0 positive roots, and there are no negative roots. Similarly, we can estimate the number of positive roots of the polynomial $p_4(x)$ (Table  \ref{Table}).
\begin{table}
  \centering
  \caption{The table of possible positive or negative roots of the polynomial $p_4(x)=A x^4-B x^3+C x^2-B x+A$ }\label{Table}
  \begin{tabular}{|c|c|c|c|c|}
    \hline
    A & B & C & Positive roots & Negative roots\\\hline
    + & + & + & 4,2,0 & 0 \\\hline
    + & + & - & 2,0 & 2,0\\\hline
    + & - & + & 2,0 & 2,0 \\\hline
    + & - & - & 0 & 4,2,0 \\\hline
    - & + & + & 2,0 & 2,0 \\\hline
    - & + & - & 0 & 4,2,0\\\hline
    - & - & + & 2,0 & 2,0 \\\hline
    - & - & - & 4,2,0 & 0 \\
    \hline
  \end{tabular}
\end{table}

Elementary analysis allows us to obtain the fixed points of function \eqref{recurrence15} by finding real positive roots of equation \eqref{recurrence16}. Thus, we can obtain a polynomial equation of degree 5. Previously documented analysis has solved these equations, which we will not show here due to the complicated nature of formulas and coefficients \cite{Wolfram}. Nonetheless, we have manipulated the polynomial equation $p_4(x)$ via Mathematica \cite{Wolfram}. 

\begin{figure} [!htbp]
\centering
\includegraphics[width=62mm]{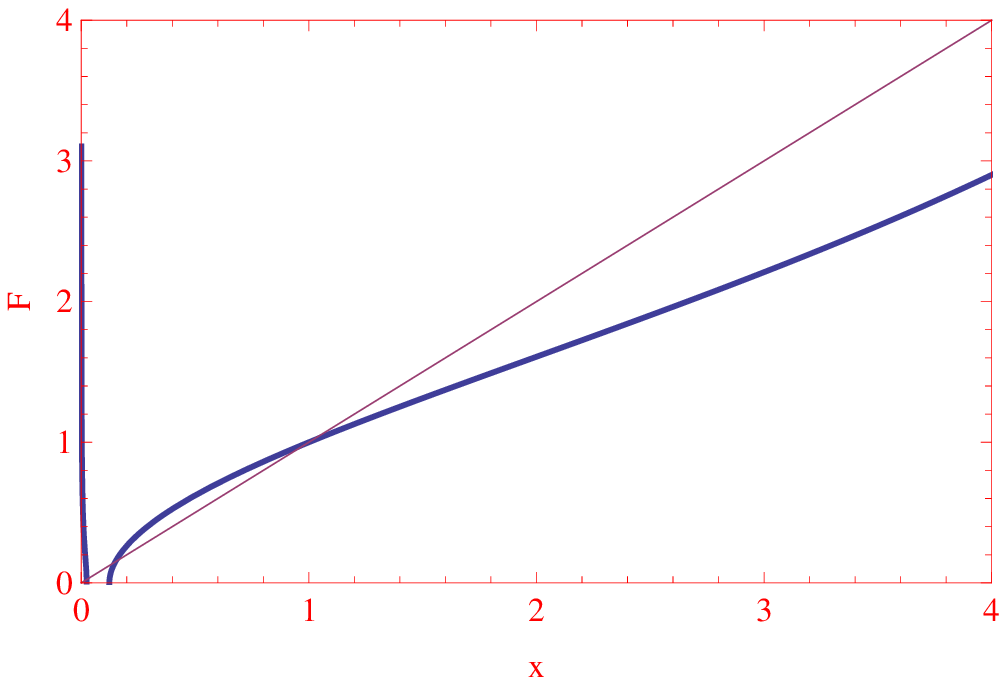}\ \ \ \ \ \ \ \
\includegraphics[width=62mm]{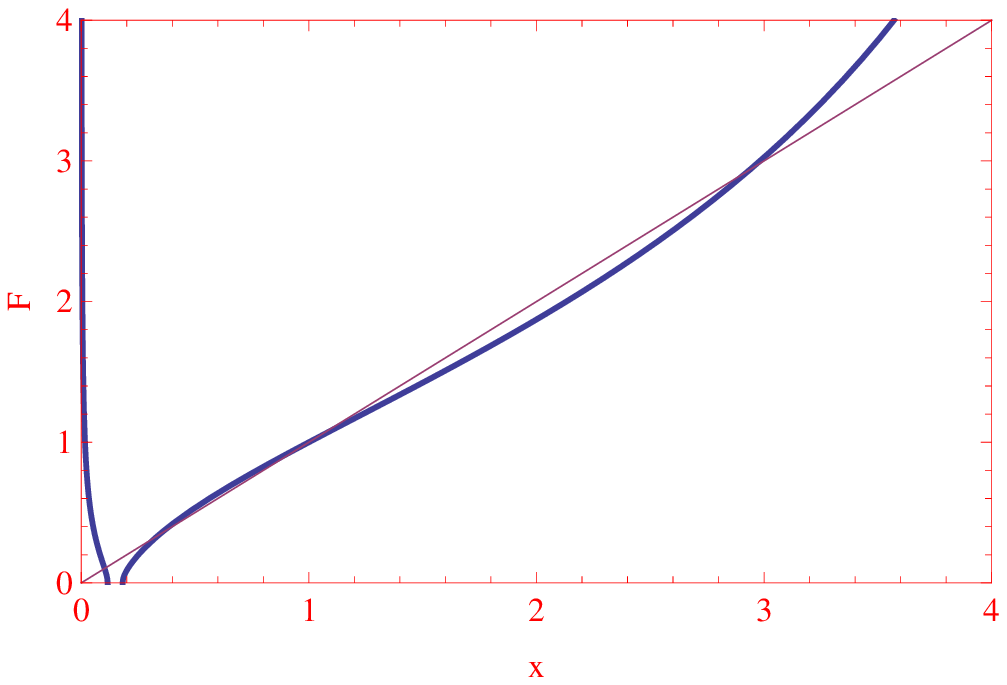}
\caption{ (a) (left) There exist  three positive roots of the equation \eqref{recurrence15} for $a = 9, b = 0.9$ and $m=1$,
(b) (right) There exist five positive roots of the equation \eqref{recurrence15} for $a = 9, b = 0.61$ and $m=1$.}
\label{fig6bnew}
\end{figure}

For the Third Case, we have obtained 5 positive real roots for some parameters $a$ and $b$. In Figure \ref{fig6bnew} (a), for $a = 9, b = 0.9$ and $m=1$, we obtained 3 positive roots. In Figure \ref{fig6bnew} (b), for $a = 9, b = 0.61$ and $m=1$, we have five positive roots. Therefore, we have demonstrated the occurrence of phase transitions in this case.

\begin{rem}
We conclude that there are at most 5 translation-invariant Gibbs measures corresponding to the positive real roots of equation \eqref{recurrence15}. Also, one can show that translation-invariant Gibbs measures corresponding stable solutions are extreme.
\end{rem}

\section{The periodic Gibbs measures}

The notion of periodic Gibbs measures is discussed by Sinai \cite{Sinai} and  Ganikhodjaev and Rozikov \cite{GR}. This section discusses periodic solutions of equations \eqref{necessary}.

To describe phases (Gibbs measures) of a given Hamiltonian on a Cayley tree, we must have correspondence between Gibbs measures and a collection of vectors $\{h_{\ell}: \ell\in L\}$, that satisfy a given non-linear equation. We denote the periodic point of the function $f$ given in \eqref{recurrence6a} as a point, $v \in \mathbb{R}_{+}$, if there exists $p$ so that $f^{p}(v) = v$ where $f^{p}$ is the $p^{th}$ iterate of $f$. The smallest positive integer $p$ satisfying the above conditions is the least period of the point $v$. We denote the set of periodic points with period $p$ with the notation $Per_p(f)$ (see \cite{AGUT,GR}).

To describe the periodic Gibbs measures, we will analyze the equation $f (f(v)) = v$, where
\begin{equation*}\label{recurrence14}
f(v):=\left(\frac{a b v^2+1}{b+a v^2}\right).
\end{equation*}
In this case, the positive roots of the equation
\begin{equation}\label{periodic1}
\frac{f(f(v))-v}{f(v)-v}=\frac{\left(b+a v^2\right) \left(a b+b^2-a( 1-b^2)v+a b(1+a b) v^2\right)}{a+b^3+2ab( a + b)v^2+a^2 b(1+a b) v^4}=0
\end{equation}
describe the periodic non transition-invariant Gibbs measures with memory of length 2.

In order to find the positive roots of equation \eqref{periodic1}, we should locate the positive roots of the following equation:
\begin{equation}\label{periodic2}
\left(a b+a^2 b^2\right) v^2-a\left(1-b^2\right) v+a b+b^2=0
\end{equation}
The discriminant of equation \eqref{periodic2} is equal to
$$
\Delta_p
= a \left(a-b^2 \left(6 a+4 b+4 a^2 b+3 a b^2\right)\right).
$$
\begin{figure} [!htbp]
\centering
\includegraphics[width=63mm]{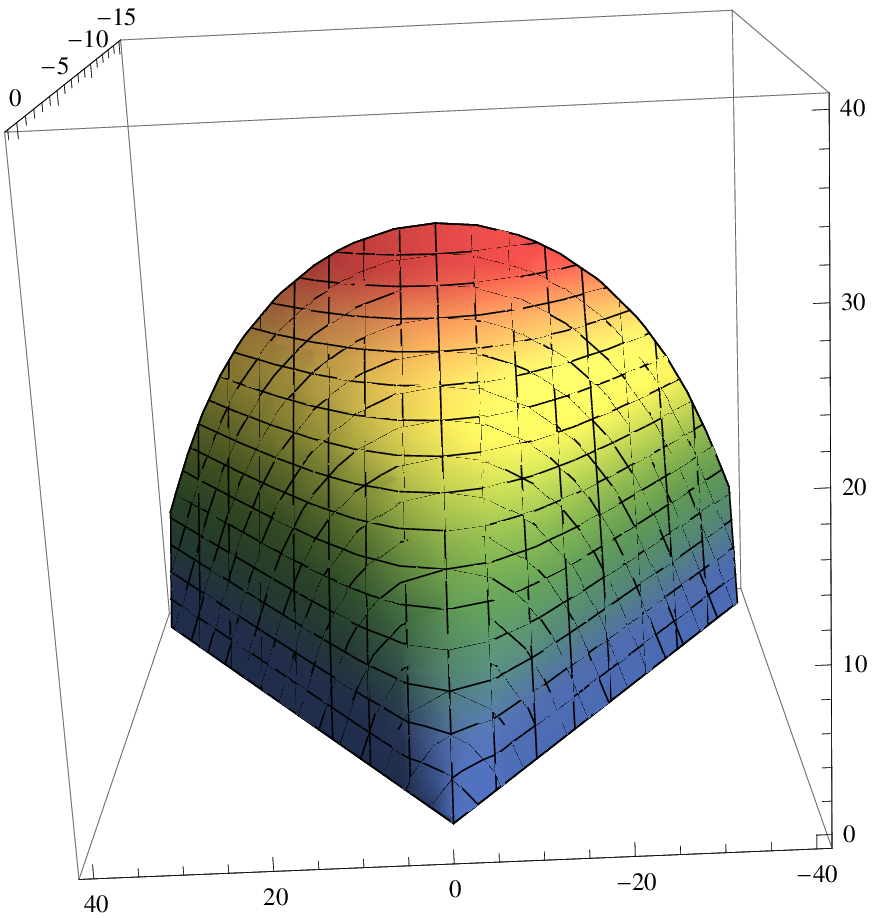}\ \ \ \ \ \ \ \ \ \
\includegraphics[width=63mm]{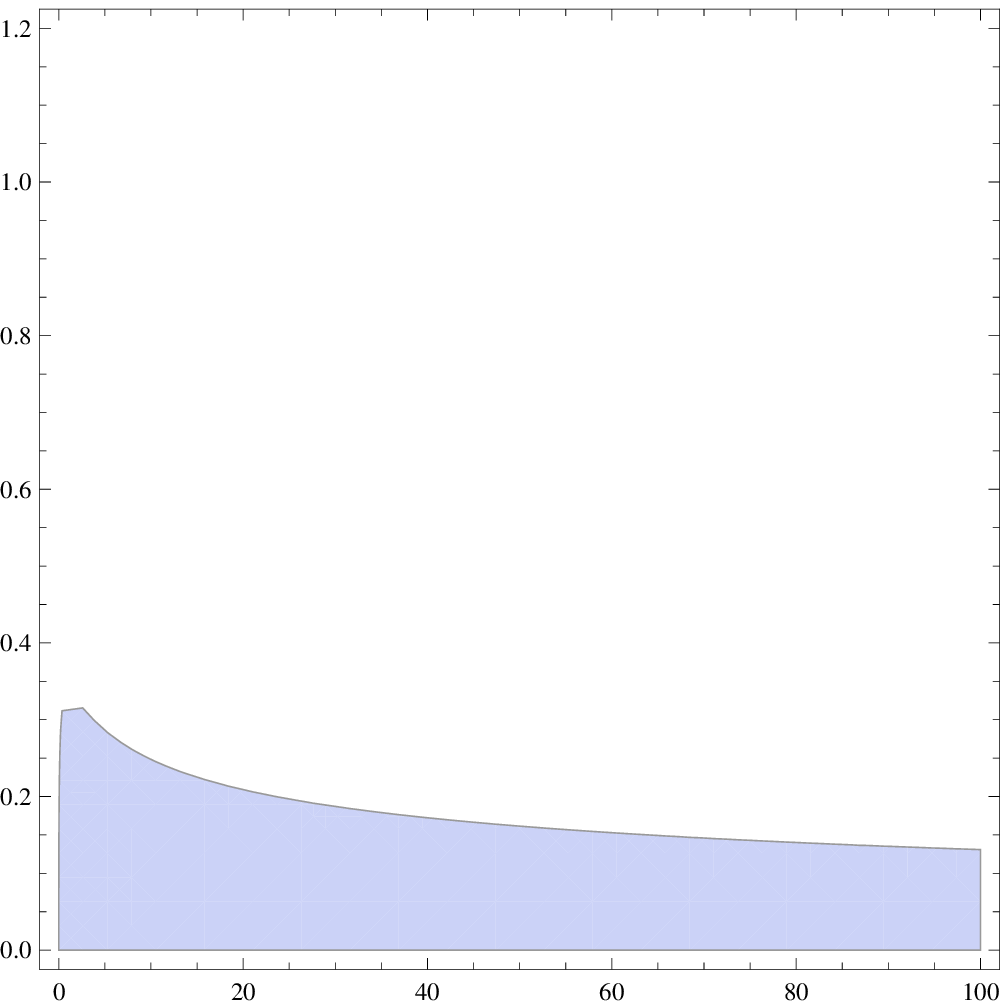}
\caption{(a) Region of periodic non transition-invariant Gibbs measures. For all $(J,J_p,T)$ inside the surface
($J\in (-40, 40), J_p\in (-15, 1), T\in(T, 0.00012, 40)$), a phase transition occurs.
(b) Region of 2 positive periodic roots with period of 2 ($a\in (0, 100), b\in (0, 1.2))$.}\label{fig3periodic}
\end{figure}
If $\Delta_p>0$ and $b<1$, then equation \eqref{periodic2} has 2 positive roots (Figure \ref{fig3periodic} (b)). Therefore, there are at most 2 periodic non transition-invariant Gibbs measures corresponding to the model \eqref{hm}. When $\ln(2J_p)<\ln(T)$, \emph{i.e.}, when there is a region of periodic non translation-invariant Gibbs measures, then for all $(J,J_p,T)$  inside the surface ($J\in (-40, 40), J_p\in (-15, 1), T\in(T, 0.00012, 40)$) there exist 2 non translation-invariant Gibbs measures (Figure \ref{fig3periodic} (a)).

\begin{thm}\label{priodicthm} For the generalized ANNNI model \eqref{hm}, in the case of non translation-invariant periodic state with period 2, a phase transition occurs for arbitrary pairs $(a,b)$ such that $\Delta_p>0$ and $b<1$.
\end{thm}
From Theorem \ref{priodicthm}, if the conditions $\Delta_p>0$ and $b<1$ are satisfied, then for the model \eqref{hm} there are two $G_2^{*}$-periodic Gibbs measures $\mu_1^{per},\mu_2^{per}$ with memory of length 2 \cite{GR}.

\begin{rem}
From Theorem \ref{priodicthm}, we have
\begin{enumerate}
  \item It follows that the measures $\mu_1^{per},\mu_2^{per}$ are non translation-invariant periodic state with period 2;
  \item In Figure \ref{fig3periodic}, for all $(J,J_p,T)$, such that $J\in (-40, 40), J_p\in (-15, 1) $ and $T\in(T, 0.00012, 40)$ inside the surface, a phase transition occurs.
\end{enumerate}
\end{rem}
\begin{thm}\label{theorem4}
All commensurate phases with finite period 2 are extreme Gibbs distributions.
\end{thm}
For the proof of \ref{theorem4}, see \cite{NHSS1,NHSS,Preston}.
\begin{rem}
Note that one may locate other periodic non transition-invariant Gibbs measures by means of the equations obtained in the Second and the Third cases.
\end{rem}
\section{Conclusions}
This paper documents several analytical points:
\begin{itemize}
  \item We have used a new approach to investigate Gibbs measures of Vannimenus model on a Cayley tree of order 2.
  \item We used the Markov random field method to describe Gibbs measures.
  \item We have constructed recurrence equations corresponding to a generalized ANNNI model.
  \item We have proved the Kolmogorov consistency condition.
  \item We have investigated the translation-invariant and periodic (non translation-invariant) Gibbs measures for the specified model.
\end{itemize}
We stress that the specified model was investigated only numerically, without rigorous mathematical proofs \cite{NHSS}. This paper has thus proposed a rigorous measure-theoretical approach to investigate Gibbs measures with memory of length 2 corresponding to the Ising-Vannimenus model on a Cayley tree of order 2. We have obtained new Gibbs measures that differ from those given in the references \cite{NHSS,FreeMA}.

\textbf{Acknowledgments} The author would like to thank Thomas Langtry for many very helpful comments.


\begin{thebibliography}{99}
\bibitem{Iof} Ioffe, D.,
A note on the extremality of the disordered state for the Ising model on the Bethe lattice
{\it Lett. Math. Phys.,} \textbf{37}, 137-143 (1996).
\end{thebibliography}
\end{document}